\documentclass[useAMS,usenatbib,usegraphicx]{mn2e}
\usepackage{amsmath,amssymb,color}

\title[KHI with GSPH]{Kelvin-Helmholtz instabilities with
{Godunov SPH}}
\author[Cha et al.]{Seung-Hoon Cha$^{\dagger}$\thanks{E-mail:
seunghoon.cha@astro.le.ac.uk}, Shu-ichiro Inutsuka$^{\ddagger}$ and
Sergei Nayakshin$^{\dagger}$\\
$^{\dagger}$Department of Physics \& Astronomy, University of Leicester,
Leicester, LE1 7RH, UK\\
$^{\ddagger}$Department of Physics, Nagoya University, Nagoya,
463--8602, Japan}
\begin{document}

\date{Accepted ???? ??? ??. Received ???? ??? ??; in original form ????
???? ??}

\pagerange{\pageref{firstpage}--\pageref{lastpage}} \pubyear{2002}

\maketitle

\label{firstpage}

\begin{abstract}
Numerical simulations for the non-linear development of Kelvin-Helmholtz
instability in two different density layers have been performed with the
particle-based method (Godunov SPH) developed by Inutsuka (2002). The
Godunov SPH can describe the Kelvin-Helmholtz instability even with a
high density contrast, while the standard SPH shows the absence of the
instability across a density gradient (Agertz et al. 2007). The
interaction of a dense blob with a hot ambient medium has been performed
also. The Godunov SPH describes the formation and evolution of the
fingers due to the combinations of Rayleigh-Taylor, Richtmyer-Meshkov,
and Kelvin-Helmholtz instabilities. The blob test result coincides well
with the results of the grid-based codes.

An inaccurate handling of a density gradient in the standard SPH has
been pointed out as the direct reason of the absence of the
instabilities. An unphysical force happens at the density gradient even
in a pressure equilibrium, and repulses particles from the initial
density discontinuity. Therefore, the initial perturbation damps, and a
gap forms at the discontinuity. The unphysical force has been studied in
terms of the consistency of a numerical scheme. Contrary to the standard
SPH, the momentum equation of the Godunov SPH doesn't use the particle
approximation, and has been derived from the kernel convolution or a new
Lagrangian function. The new Lagrangian function used in the Godunov SPH
is more analogous to the real Lagrangian function for continuum.
The momentum equation of the Godunov SPH has much better linear
consistency, so the unphysical force is greatly reduced compared to the
standard SPH in a high density contrast.

\end{abstract}

\begin{keywords}
hydrodynamics -- instabilities -- turbulence -- methods: numerical --
galaxies: formation -- galaxies: evolution -- star: formation
\end{keywords}

\section{Introduction}
SPH \citep[Smoothed Particle Hydrodynamics,][]{Gingold1977a,Lucy1977a}
is a fully lagrangian and gridless method, and has been used widely
in the various fields of astrophysics \citep{Monaghan1992a},
especially, in an irregular--shaped and/or
self-gravitating system. It is because of its lagrangian nature
and also due to the incorporation of the tree-structure \citep{Barnes1986a}.
The tree structure is very efficient not only in the calculation of the
gravity, but also in finding neighbours. Therefore, SPH becomes a very
effective tool in the research of star or galaxy formation.

However, \citet{Agertz2007a} (hereafter A07) showed that SPH has a
difficulty to describe the Kelvin--Helmholtz instability (hereafter KHI)
across a density gradient. They performed KHI simulations in
the two different density layers with two standard SPH codes
(GADGET2 \citep{Springel2001a},
GASOLINE \citep{Wadsley2004a})
and five grid--based codes
(ART \citep{Kravtsov1997a},
CHARM \citep{Miniati2007a},
ENZO--PPM \citep{Bryan1997a},
ENZO--ZEUS \citep{Stone1992a},
FLASH \citep{Fryxell2000a}).
A complete absence of KHI across a density gradient has been observed in
the results of the standard SPH codes. However, there are nicely rolled
vortices in the simulations with the grid--based codes even in a high
density contrast. The standard SPH codes show the vortices in the 
homogeneous density case only. They also
performed the interaction of a blob and a hot ambient medium with
a high mach number (the blob test).
In the results of the grid--based codes,
fingers are initiated due to the Rayleigh--Taylor and
Richtmyer--Meshkov instabilities at the front of the compressed blob,
and then
enhanced by the KHI. Finally, the blob is destroyed. However,
the standard SPH codes show only compression of the blob.
They called it ``the fundamental difference'' between
the standard SPH and the grid--based
codes. Their results should be a big problem, because the KHI plays
an important role in the various fields where SPH
has been applied intensively.

A07 found that there is a strange behaviour of particles
around the initial contact discontinuity of the two different density
layers. Particle alignments are observed, and a gap forms
along the initial contact discontinuity.
The gap formation or ``the peeling'' of the particle layers around a
density gradient has been reported already \citep{Fulk1994a},
and the modification of the initial particle
configuration has been suggested as a prescription
\citep{Monaghan1987a, Fulk1994a}. Therefore,
A07 performed the same simulation with three different
initial conditions, such as the lattice, poisson and glass.
Although the different initial conditions showed
different results to each other,
the KHI was still absent in the standard SPH simulations.
Another possibility as the reason of the gap formation is the artificial
viscosity, because the artificial viscosity of the standard SPH
has a long history of criticism \citep{Watkins1996a,Cha2003a}
in various aspects.
Although the artificial viscosity can give a minor change to the results,
the KHI is still absent irrespective
of the artificial viscosity.
Furthermore, the ENZO--ZEUS code used in A07 employs
a von Neumann--Richtmyer type artificial viscosity also, but
showed the KHI in a density gradient.
Therefore, the artificial viscosity cannot be a reason of the
gap formation. Finally, A07 concluded
that the absence of the KHI and the gap formation are
due to an inaccurate handling of
hydrodynamic force across a density gradient.

The absence of KHI in a density gradient is a serious
problem to SPH users, so there should be a quick response.
\citet{Price2008a} suggested an artificial conduction term in the
energy equation of the standard SPH.
Similar to the role of the artificial viscosity at the momentum
discontinuity,
the artificial conduction acts on the thermal energy discontinuity,
and changes the pressure profile to a continuous one
across a density gradient.
He showed that the new energy equation containing the artificial
conduction term can describe the KHI across a density gradient.
\citet{Price2008a} also expected the new
formulation of \citet[][hereafter I02]{Inutsuka2002a}
based on the new Lagrangian function may handle the density gradient
correctly.

In this paper, we will revisit the Godunov SPH (hereafter GSPH)
proposed by I02 as a possible solution of the inaccurate handling of
a density gradient.
The same tests in A07 have been performed here again
with a two--dimensional GSPH code.
The unphysical force across a density gradient
is much reduced in the GSPH results, and the KHI and other
instabilities are observed.
Especially, the KHI developing in the diagonal
direction has been simulated, and a satisfying result is obtained.
Complicated patterns due to the combinations of the 
instabilities develop in the blob test.

The inaccurate handling of a density
gradient in the standard SPH has been studied in terms of the
consistency of a numerical method, and is given in section 2. As a
prescription, the momentum equation of GSPH has been
revisited with the kernel convolution and also
the new Lagrangian function of a particle system in section 3.
The consistency of GSPH has been investigated, and a simple test to
verify the density discontinuity handling has been performed in the same
section. The KHI simulations in the two different density layers
and the blob test
are in section 4. Finally, the summary is given in section 5.
 
\section{Consistency of SPH}
\subsection{Stability, consistency and convergence}
Probably, the most important property of a numerical scheme is the
convergence, because the convergence addresses how close a numerical
solution is to the actual solution. However, it is not easy in general
to prove the convergence of a numerical scheme directly, because the
actual solution is not unveiled in most problems. Therefore, the Lax
equivalence (or Lax--Richtmyer) theorem
\citep[e.g.][]{Gary1966a,Ritchmyer1967a,Despres2003a}
is very useful to check
the convergence of a numerical scheme. According to the Lax equivalence
theorem, the stability and consistency are sufficient conditions of
the convergence.

First of all, the stability can be defined clearly, and has been
studied intensively in the standard SPH so far
\citep{Monaghan1989a,Balsara1995a,Swegle1995a,Morris1996a,Cha2003b}
at least in the linear regime. The standard
SPH is conditionally stable,
so with the CFL condition \citep{Courant1948a},
the stability of the standard SPH is
guaranteed

Secondly, the consistency of a numerical scheme means
how well the numerical equations of the
scheme approximate the physical equations
\citep{Fulk1994a,LeVeque2002a}, and is directly related to the analysis
of the truncation error. The truncation error of a numerical scheme
should vanish as
the time step, $\Delta t$ and the grid size, $\Delta x$ (in SPH, the
smoothing length, $h$ is comparable to $\Delta x$ of
grid--based codes) approach to the infinitesimal value if the
scheme has the consistency. Therefore, the loss of consistency
will lead to low accuracy of the numerical scheme. One may concentrate
on the consistency to get the convergence of the standard SPH, because
the stability is already proved.

Although the consistency problem of the standard SPH is well known
already \citep[e.g.][]{Fulk1994a,Dilts1999a},
it will be reviewed briefly in the
following sections for the convenience of readers.
Two approximations are needed to get the momentum
equation of the standard SPH. One is the kernel approximation
and the other is the particle approximation.
The consistency of the standard SPH will be examined in both of the two
approximations.

\subsection{Kernel approximation}
The kernel approximation is given by
\begin{equation}
\langle f\rangle(\boldsymbol x) =
\int f(\boldsymbol x')W(\boldsymbol x - \boldsymbol x',h)d\boldsymbol x',
\label{kernel_approxi}
\end{equation}
where $W$ and $\langle f\rangle$
are the kernel and kernel--smoothed functions, respectively.

In order to check the (order of) consistency of the kernel approximation,
we will follow the procedure of \citet{Liu2006a}.
If a numerical scheme can produce a polynomial of up to $n^{th}$--order
exactly, the numerical scheme is said to
have the $n^{th}$--order consistency.
For example, to check the $0^{th}$--order consistency of the kernel
approximation, put a constant function, $f(\boldsymbol x') = C_o$ into
Eq. (\ref{kernel_approxi}), then the kernel--smoothed function becomes
\begin{equation}
\langle f\rangle(\boldsymbol x)
= \int C_o W(\boldsymbol x - \boldsymbol x',h) d\boldsymbol x' = C_o.
\label{kernel_approxi_0}
\end{equation}
Here, the normalisation condition of the kernel function,
\begin{equation}
\int W(\boldsymbol x - \boldsymbol x',h)d\boldsymbol x' = 1
\label{nor_cond}
\end{equation}
is used. The $0^{th}$--order of consistency is easily proved by
Eq. (\ref{kernel_approxi_0}).
For the $1^{st}$--order consistency, one may put a linear function,
$C_0 + C_1\boldsymbol x'$ into $f(\boldsymbol x')$, then
\begin{eqnarray}
\nonumber
&&\langle f\rangle(\boldsymbol x) = \\
&&\int \left(C_0+C_1\boldsymbol x'\right)
W(\boldsymbol x - \boldsymbol x',h)d\boldsymbol x'
=C_0 + C_1\boldsymbol x.
\label{kernel_approxi_1}
\end{eqnarray}
Here, the normalisation condition (Eq. (\ref{nor_cond}))
and the symmetry property of the kernel function,
\begin{equation}
\int \boldsymbol x' W(\boldsymbol x - \boldsymbol x',h)
d\boldsymbol x' = \boldsymbol x
\label{nor_id1}
\end{equation}
are used.

For a higher order of consistency, we need a higher moment of a kernel
function. For example, the $2^{nd}$--order of consistency
needs that the $2^{nd}$ moment of a kernel function should be
\begin{equation}
\int \left(\boldsymbol x - \boldsymbol x'\right)^2
W(\boldsymbol x - \boldsymbol x',h)d\boldsymbol x' = 0.
\end{equation}
However, it is impossible to achieve this with a
non--negative kernel function.
Therefore, the order of consistency
of the kernel approximation is less than 2 with a non--negative and
normalised symmetric kernel function.

For a more complete discussion, we may have to repeat the same
analysis above for the first derivative of $f(\boldsymbol x)$ also,
because the hydrodynamic equations contain the first derivative of
physical quantities. However, it is not necessary to check the
consistency only, so omitted here. See \citet{Monaghan1982a} or
\citet{Liu2003a} for the further details.

The consistency depends on the
normalisation condition of the kernel.
Therefore, the kernel
approximation loses its $0^{th}$--order consistency if 
the normalisation condition
is not satisfied, for example, at the
edge of a dense cloud in a rarefied ambient medium.
However, the tests performed in A07 and also in 
Sec. \ref{sec_tests} have correct boundary treatments,
so the incompleteness
of a kernel function at the boundary is not a critical problem in the
tests.

Although the kernel approximation has the consistency, it is not directly
used in the equations of SPH. Instead of the kernel approximation, the
particle approximation is used for the derivation of the standard SPH
equations, and is explained below.

\subsection{Particle approximation}
\label{sph_pa}
The particle approximation used in the standard SPH is given by
\begin{equation}
f^{SPH}(\boldsymbol x) = 
\sum_j \frac{m_j}{\rho(\boldsymbol x_j)}f(\boldsymbol x_j)
W_j,
\label{particle_approxi}
\end{equation}
where 
$\displaystyle W_j$ is
$\displaystyle W(\boldsymbol x-\boldsymbol x_j,h)$.

We will repeat the same procedure performed in the previous section
to check the consistency of the particle approximation.
For the $0^{th}$--order, put a constant $C_o$ instead of
$f(\boldsymbol x_j)$ of Eq. (\ref{particle_approxi}), then
\begin{equation}
f^{SPH}(\boldsymbol x) = C_o\sum_j \frac{m_j}{\rho(\boldsymbol x_j)} 
W_j.
\label{particle_approxi_0}
\end{equation}
The particle approximation can reproduce the constant function when
\begin{equation}
\sum_j \frac{m_j}{\rho(\boldsymbol x_j)} W_j = 1.
\label{sph_id}
\end{equation}
Eq. (\ref{sph_id}) holds only in an even distribution of particles.
Therefore, the particle approximation loses its $0^{th}$--order
consistency in an uneven distribution of particles, and eventually
the standard SPH is unable to converge to the actual solution
in that situation.

\begin{figure}
\includegraphics[width=8.5cm]{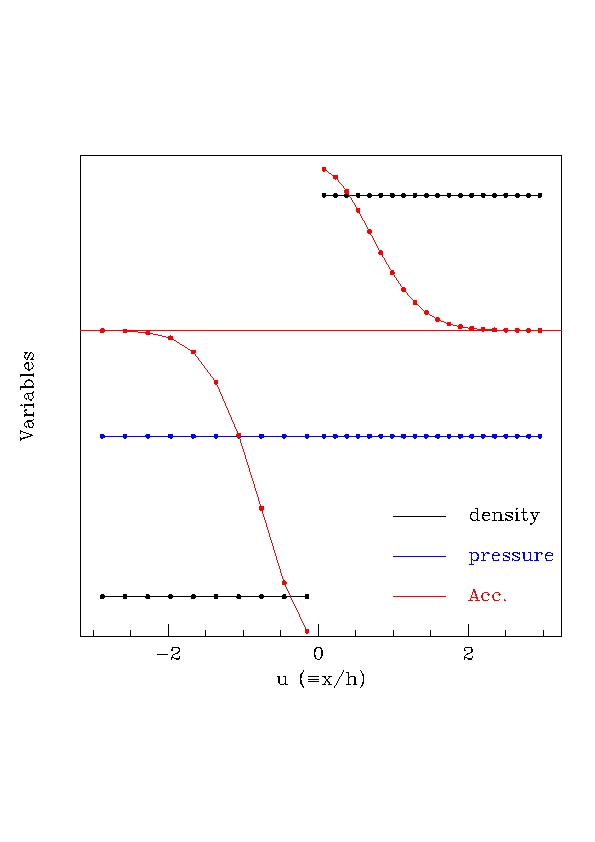}
\caption{The black and blue solid lines are density and pressure
profiles, respectively. The dots on the lines denote the particle
positions. They implement a density gradient with a
pressure equilibrium. The particle distance from the centre,
$x$ is scaled by the
smoothing length, $h$. The red solid line with dots is the acceleration
of particles calculated by the standard SPH. The red solid line without
dots shows the expected acceleration under a pressure equilibrium. A
repulsion happens at the density discontinuity, and
will damp the initial perturbation. Therefore, any instability across
the density gradient may be suppressed in the standard SPH.}
\label{sph_acc}
\end{figure}

This problem appears in the equation of motion of the standard
SPH. With a pressure equilibrium, one of the typical motion equation
of the standard SPH without the artificial viscosity may be written by
\begin{equation}
\label{motion_sph}
\boldsymbol a_i \equiv \frac{d\boldsymbol v_i}{dt} = -P
\sum_j m_j\left(\frac{1}{\rho_i^2}+\frac{1}{\rho_j^2}\right)
\frac{\partial}{\partial \boldsymbol x_i} W_{ij},
\end{equation}
where
$\displaystyle W_{ij}$ is
$\displaystyle W(\boldsymbol x_i-\boldsymbol x_j,h)$, and
the physical variables have their usual meaning.
Although a pressure equilibrium is assumed,
the hydrodynamic acceleration of particle $i$
doesn't vanish, so the particle will move.
Figure \ref{sph_acc} shows this unphysical force. The calculated
acceleration (red solid line with dots)
should vanish because the pressure is constant across the density
discontinuity. However, the acceleration shows a
repulsion of particles at the discontinuity. This repulsion damps
the initial perturbation, and suppresses the KHI. It will make a gap
between the two different density layers as well.
Therefore, one may understand that the occurrence of the
unphysical force across a density gradient in a pressure equilibrium is
due to the loss of the $0^{th}$--order consistency.

The only way to eliminate the unphysical force is to make
the density term inside the round brackets of Eq. (\ref{motion_sph}) an
even function.
Especially, a uniform density field around particle $i$ is
the interesting case, and this is why the KHI appears in the homogeneous
density case (1:1 density contrast case) in A07. However, a uniform
distribution of particles is a special situation, not a general one.

\section{Consistency of Godunov SPH}
\subsection{Kernel convolution}
The inconsistency of the standard SPH is due to
Eq. (\ref{sph_id}), which appears
in the conversion from a continuum (the kernel approximation)
to a particle system (the particle approximation).
I02 and \citet{Dilts1999a} pointed out Eq. (\ref{sph_id}) as a crude
assumption, and I02 suggested a density estimation at
a arbitrary position $\boldsymbol x$,
\begin{equation}
\rho(\boldsymbol x) \equiv \sum_j m_j W_j.
\label{density_eq}
\end{equation}
With Eq. (\ref{density_eq}), two identities,
\begin{equation}
1 = \sum_j \frac{m_j}{\rho(\boldsymbol x)}W_j
\label{id1}
\end{equation}
and
\begin{equation}
0 = \sum_j m_j\frac{\partial}{\partial \boldsymbol x}
\frac{W_j}{\rho(\boldsymbol x)}
\label{id2}
\end{equation}
are derived.
Instead of Eq. (\ref{sph_id}),
Eq. (\ref{id1}) has been directly cast into the kernel approximation
(Eq. (\ref{kernel_approxi})) to avoid the
inconsistency of the particle approximation.
Therefore, the final form of the kernel approximation of GSPH becomes
\begin{eqnarray}
\nonumber
&&\langle f\rangle(\boldsymbol x) = f^{GSPH} (\boldsymbol x)\\
&=& \int \sum_j m_j\frac{f(\boldsymbol x')}{\rho(\boldsymbol x')}
W(\boldsymbol x'-\boldsymbol x,h)W(\boldsymbol x'-\boldsymbol x_j,h)
d\boldsymbol x'.
\label{new_kernel_approxi}
\end{eqnarray}
The meaning of Eq. (\ref{new_kernel_approxi}) is clear. GSPH considers 
both host and neighbour particles as an extended particle, and uses the
information of the detailed internal
structure of the two extended particles.
Note that the standard SPH considers
the host particle as a smoothed one by the contributions of its neighbours,
but the neighbour particles are still a point.
$f^{GSPH} (\boldsymbol x)$ of Eq. (\ref{new_kernel_approxi})
reduces to the particle approximated function
if particle $j$ is considered as a point.
If $W(\boldsymbol x'-\boldsymbol x_j,h)$ of
Eq. (\ref{new_kernel_approxi})
is approximated by the delta function, $\delta(\boldsymbol x'
- \boldsymbol x_j)$, then Eq. (\ref{new_kernel_approxi}) becomes
\begin{equation}
f^{GSPH} (\boldsymbol x) =
\sum_j \frac{m_j}{\rho(\boldsymbol x_j)}f(\boldsymbol x_j)W_{j},
\end{equation}
which is identical to Eq. (\ref{particle_approxi}) if it is evaluated
at $\boldsymbol x = \boldsymbol x_i$.

In order to check the consistency of GSPH,
put a linear function,
$C_0 + C_1\boldsymbol x'$ into $f(\boldsymbol x')$
of the right--hand--side of Eq. (\ref{new_kernel_approxi}),
then

\begin{eqnarray}
\nonumber
&&f^{GSPH} (\boldsymbol x)
= \int \sum_j\frac{m_j}{\rho(\boldsymbol x')}\\
\nonumber
&\cdot&\left(C_0+C_1\boldsymbol x'\right)
W(\boldsymbol x'-\boldsymbol x,h)W(\boldsymbol x'-\boldsymbol x_j,h)
d\boldsymbol x'\\
&=& \int \left(C_0+C_1\boldsymbol x'\right)\\
\nonumber
&\cdot&
\left[\sum_j\frac{m_j}{\rho(\boldsymbol x')}
W(\boldsymbol x'-\boldsymbol x_j,h)\right]
W(\boldsymbol x'-\boldsymbol x,h) d\boldsymbol x'\\
\nonumber
&=& C_0 + C_1\boldsymbol x,
\end{eqnarray}
where Eqs. (\ref{nor_cond}), (\ref{nor_id1}) and (\ref{id1}) are used.
Eq. (\ref{new_kernel_approxi})
can reproduce the linear function, so the first order of
consistency is guaranteed in GSPH.

Finally, the momentum equation of GSPH is derived
using Eqs. (\ref{density_eq})
- (\ref{new_kernel_approxi}), and becomes
\begin{equation}
\label{motion_gsph}
\frac{d\boldsymbol v_i}{dt} = 
-\sum_j m_j\int\frac{P(\boldsymbol x)}{\rho^2(\boldsymbol x)}
\left(\partial_i-\partial_j\right)W_iW_j d\boldsymbol x,
\end{equation}
where
$\displaystyle W_i$,
$\partial_i$ and $\partial_i$ are 
$\displaystyle W(\boldsymbol x-\boldsymbol x_i,h)$,
$\frac{\partial}{\partial \boldsymbol x_i}$ and
$\frac{\partial}{\partial\boldsymbol x_j}$, respectively.
Contrary to the momentum equation of the standard SPH,
Eq. (\ref{motion_gsph})
is expected to converge to the actual solution
even in a large density gradient.
We investigate the behaviour of Eq. (\ref{motion_gsph})
at a general density field with a pressure equilibrium to check this.
\begin{eqnarray}
\nonumber
\frac{d\boldsymbol v_i}{dt} &=& -P\int\sum_j m_j
\frac{1}{\rho^2(\boldsymbol x)}\left(\partial_i-\partial_j\right)
W_iW_jd\boldsymbol x\\
\nonumber
&=& -P\int\sum_jm_j
\left[ \frac{W_j}{\rho(\boldsymbol x)}
\frac{\partial_i W_i}{\rho(\boldsymbol x)}
- \frac{W_i}{\rho(\boldsymbol x)}
\frac{\partial_jW_j}{\rho(\boldsymbol x)}\right]
d\boldsymbol x\\
&=& -P\int\left[
\frac{\partial_iW_i}{\rho(\boldsymbol x)}
+\frac{W_i}{\rho^2(\boldsymbol x)}\frac{\partial \rho(\boldsymbol x)}{\partial \boldsymbol x}\right]d\boldsymbol x\\
\nonumber
&=& -P\int \frac{\partial}{\partial \boldsymbol
x}\left[\frac{W_i}{\rho(\boldsymbol x)}\right]d\boldsymbol x\\
\nonumber
&=& 0,
\end{eqnarray}
where Eqs. (\ref{id1}) and (\ref{id2}) are used.
One can see the acceleration calculated by Eq. (\ref{motion_gsph})
vanishes in the pressure equilibrium regardless of the density field,
to the degree to which the interpolation of the density field used to
compute the integral in GSPH (see I02) is exact.

\subsection{Perturbation damping test}
\begin{figure}
\includegraphics[width=8.5cm]{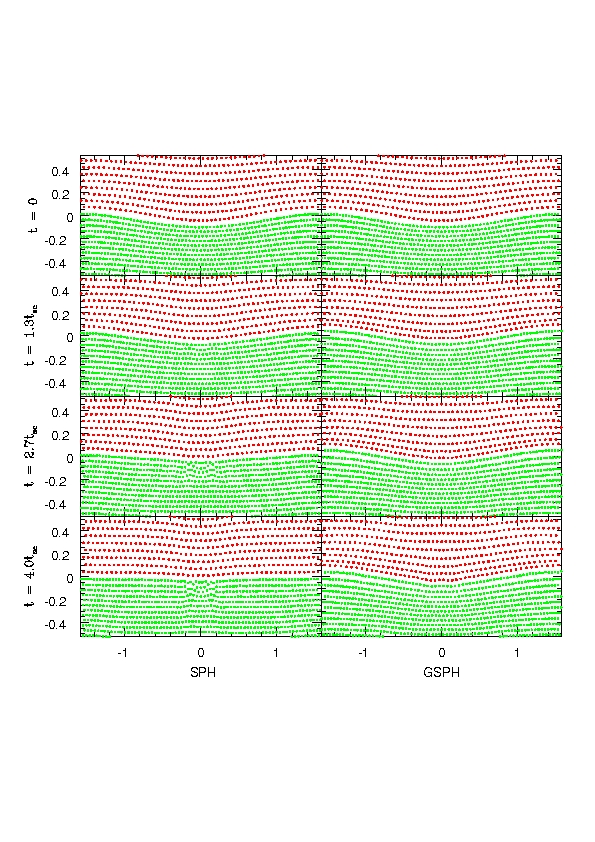}
\caption{
The perturbation damping test with the standard SPH (right column)
and GSPH (right column).
Individual snapshot shows the particles position at 
$\displaystyle t=0.0,\,1.3,\,2.7\text{ and }4.0t_{sc}$ from
the top, respectively.
In the standard SPH results, the initial perturbation is
damped completely in $\displaystyle 4t_{sc}$ due to the repulsion
of particles across the density gradient, although the pressure is
uniform. In contrast, the perturbation
survives in the GSPH results. The initial density contrast is 1:2
in the upper and lower layers.
}
\label{pos_damping}
\end{figure}

In order to observe the particle behaviour in the standard SPH and GSPH
at a density gradient, a test for the
damping of a perturbation has been performed.
A two--dimensional calculation domain,
$\displaystyle \left[-L_x,L_x\right] \times
\left[-L_y,L_y\right]$ has been set. Here, $L_x = L_y = \pi/2$.
Particles are located in a lattice by the $\Delta x/2$ offset initially,
and the density contrast is $1:2$ between upper and lower layers. 
A small displacement of position, $\boldsymbol\xi (\xi_x,\xi_y)$,
\begin{eqnarray}
&\begin{cases}
\xi_x=-A\kappa\sin(\kappa x)[\exp(-\kappa y)+\exp(\kappa y - 2\kappa
L_y)]\\
\xi_y=-A\kappa\cos(\kappa x)[\exp(-\kappa y)-\exp(\kappa y - 2\kappa
L_y)],
\end{cases}
\\
&\begin{cases}
\xi_x=-A\kappa\sin(\kappa x)[-\exp(\kappa y)-\exp(-\kappa y - 2\kappa
L_y)]\\
\xi_y=A\kappa\cos(\kappa x)[-\exp(\kappa y)+\exp(-\kappa y - 2\kappa
L_y)],
\end{cases}
\end{eqnarray}
is added to the upper and lower layers, respectively.
Here, $x$ and $y$ are the initial position of the particles (i.e. the
lattice). The amplitude and wavenumber of the position displacement,
$A$, $\kappa$ are set to $0.01\pi$ and $2$, respectively.
$\boldsymbol\xi$ is added to the original position to move the particle
to the perturbed position. A pressure equilibrium is assumed
in the whole calculation domain, and the sound speed of the upper layer
is set to 1. The sound crossing time, $t_{sc}$ of the vertical
direction in the upper layer
becomes $\pi/2$. Figure \ref{pos_damping} shows the snapshots of the
standard SPH and GSPH results at the different times.

Any movement of the particles is not expected in the test, because the
pressure is uniform in the whole calculation domain. However, the
initial contact discontinuity becomes flattened in the standard SPH
results. The contact discontinuity of the GSPH results preserves its
initial shape well. It is clear that the repulsion due to the
inconsistency of the standard SPH damps the
perturbation. The repulsive force acts on the normal direction
with respect to the
density discontinuity, so seems to be a surface tension
\citep{Price2008a}.
We have changed the curvature of the initial perturbation,
and confirm that the damping depends on the curvature. 

\subsection{Lagrangian function}
Another way to derive the equations of SPH is the use of a Lagrangian
function. \citep[e.g.][and references therein]{Price2004b}.
The Lagrangian function, $L$ of a fluid is given by
\begin{equation}
L = \int\rho\left(\frac{1}{2}\boldsymbol v^2 - u\right)d\boldsymbol x,
\end{equation}
where $u$ is the specific internal energy. 
The Lagrangian function of the standard SPH is
\begin{equation}
L_{SPH} = \sum_i m_i \left(\frac{1}{2} \dot{\boldsymbol x}_i^2- u_i\right),
\label{l_of_sph}
\end{equation}
With this Lagrangian function, the Euler--Lagrange equation gives
the motion equation of the standard SPH.
However, the particle approximation is already used in the Lagrangian
function (Eq. (\ref{l_of_sph})), so the resulting momentum equation
from the
Lagrangian function still has the inconsistency in the uneven particle
distribution.

The relation between the Lagrangian function and the exact
fluid Lagrangian function is shown in I02.
He derived the exact Lagrangian function of a particle system, and then
make an approximated Lagrangian,
\begin{equation}
L_{NEW} = \sum_i m_i \left[\frac{1}{2}\dot{\boldsymbol x}_i^2-
\int u({\boldsymbol x})W_i d{\boldsymbol x}\right],
\label{l_of_gph}
\end{equation}
 which has the $2^{nd}$--order accuracy.
The new Lagrangian function is very similar to that of the standard SPH,
but the only difference is the specific internal energy term.
The specific internal energy appears as if
smoothed once more than the standard SPH,
but this form as the second term in the Lagrangian function
is exactly the same as the corresponding term in the Lagrangian
function for real fluid (see Eqs. (29) and (41) of I02).
The momentum equation derived by the use of Eq.
(\ref{l_of_gph}) is the exactly same as the equation derived by the
kernel convolution.

In order to integrate Eq. (\ref{motion_gsph}),
functional forms of the density and pressure are needed.
The linear or cubic spline interpolation has been used in I02
as the function of the density around the particles $i$ and $j$,
but there is a room for the further improvement
for a more accurate handling of the density field.
For the determination of the pressure and velocity
between the particles $i$ and $j$, a riemann problem solver
(hereafter RPS) has been used.
This is why this method is called the ``Godunov SPH''. As the
usual Godunov grid--based method, any kind of
explicit
dissipation (e.g. artificial viscosity)
is not needed by the virtue of the RPS.

Note that the use of an RPS in GSPH has no direct relation to either the
absence of the KHI or the consistency problem. The unphysical force due
to the inconsistent momentum equation of the standard SPH has been
fixed by the new momentum equation of GSPH derived from
the kernel convolution or the new Lagrangian function.
The RPS is used for the
description of shock waves, because it generates a small but sufficient
numerical dissipation around shock waves. In order to check this point,
we have performed the KHI simulations with
the simplest version of GSPH suggested by \citet{Cha2003b}.
The simplest GSPH uses the same momentum equation of the standard SPH,
but employs an RPS instead of the artificial viscosity.
The simplest GSPH shows also the absence of the KHI in a density
gradient.

\section{Tests}
\label{sec_tests}
Two kinds of test have been performed. One is the traditional KHI
simulation in the two layers with a velocity shear, and the other is
the blob test. All tests have been performed with a two--dimensional
$2^{nd}$--order\footnote{The KHI can be triggered with the $1^{st}$--order
scheme, but doesn't develop very well.
The details to implement the
$2^{nd}$--order GSPH scheme is very similar to the MUSCL scheme
\citep{vanLeer1997a}, and will be omitted because it is described in
I02.}
GSPH code
incorporated with the adiabatic equation of state. The specific heat ratio,
$\gamma$ is set to $5/3$ in all simulations.

\subsection{KHI in the two-layers $(\rho_{u}:\rho_{l} = 1:2)$}
\label{sec_2khi}
There are two layers with the different density in a pressure
equilibrium initially. The equilibrium pressure is set to $2.5$ in code
unit, and the density ratio between the upper and lower layers
is set to 1:2. The two layers move to the opposite direction to each
other with the mach numbers $0.22$ and
$0.3$ in the upper and lower layers, respectively.
The whole calculation domain is
$\left[0,\frac{1}{3}\right]\times\left[-\frac{1}{6},\frac{1}{6}\right]$.
The size of calculation domain is smaller than that of A07
in order to save the calculation time.
The periodic and mirror boundary conditions have been implemented
in the $x$ and $y$--directions, respectively.
The total number of particles inside the calculation domain is $\simeq
10^5$, and the initial configuration of the particle distribution
is the lattice (A07).

An initial velocity perturbation in the $y$--direction is given by
\begin{equation}
A_o \sin\left(\frac{2\pi x}{\lambda}\right),
\end{equation}
where $A_o$ is the amplitude of the perturbation, and set to $1/40$ of
the initial velocity shear.
Here, $\lambda$ is the wavelength of
the initial perturbation, and is set to $1/6$. Therefore, two vortices
are expected in the calculation domain.
The initial perturbation is given only in a thin layer
($|y| < 0.05$) around the initial contact discontinuity

With the initial velocity shear and the density contrast, the KHI time
scale is defined by
\begin{equation}
\tau_{KH} = 
\frac{\lambda(\rho_{u}+\rho_{l})}{v_{shear}\sqrt{\rho_{u}\rho_{l}}}.
\label{tkh}
\end{equation}
Here
$\rho_{u}$ and $\rho_{l}$ are the densities of the upper and lower layers,
respectively, and $v_{shear}$ is the velocity difference between the
two layers. $\tau_{KH}$ is 0.43 in code unit.

\begin{figure*}
\includegraphics[width=8cm]{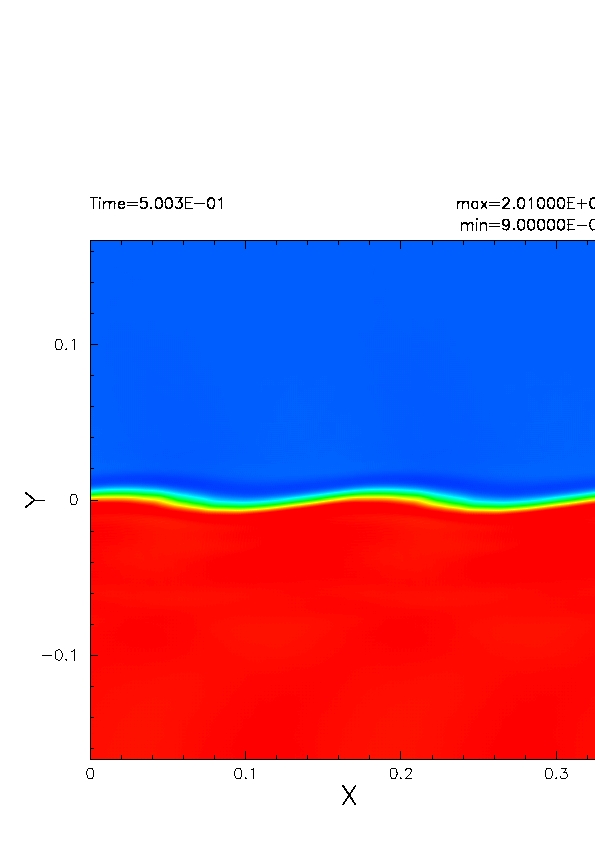}
\includegraphics[width=8cm]{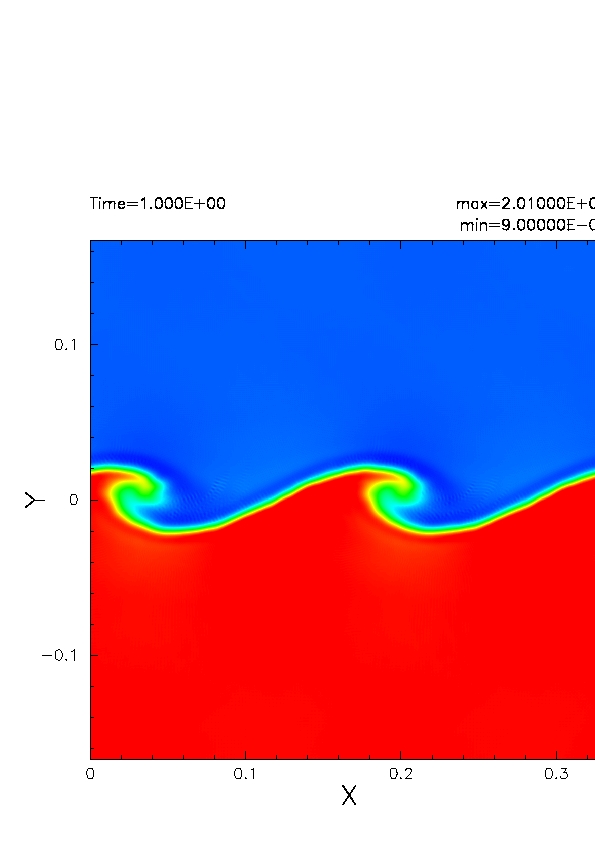}
\includegraphics[width=8cm]{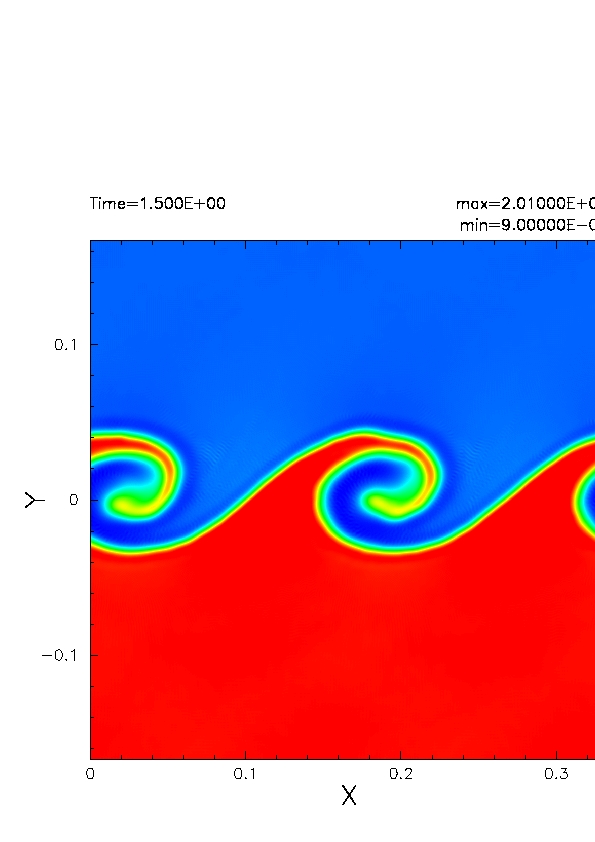}
\includegraphics[width=8cm]{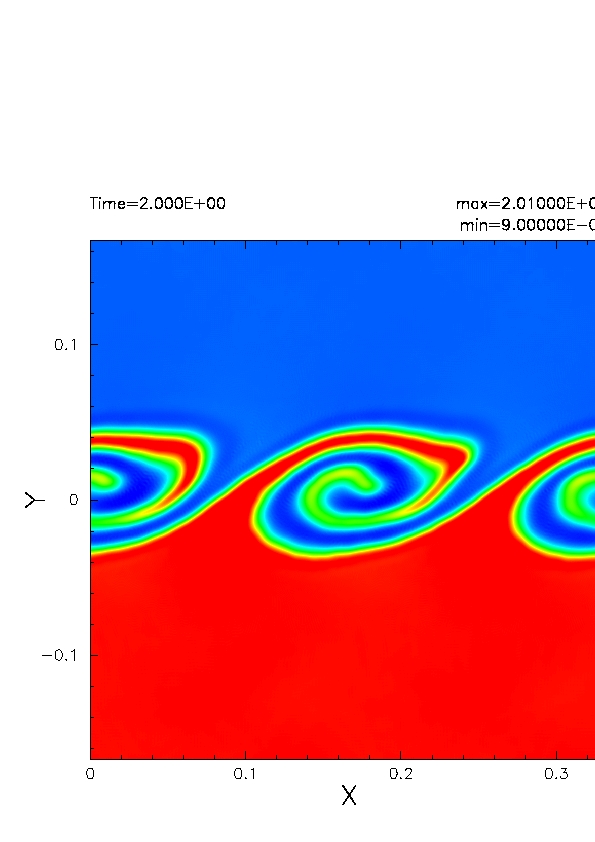}
\caption{The KHI simulation in the two different density layers.
The initial density contrast between the layers is 1:2,
and the initial mach numbers of the upper and lower layers are set to
$0.22$
and $0.3$, respectively. The upper layer moves to the right and the
lower layer moves to the left. The initial contact discontinuity
between the two layers begins wiggling due to the
initial perturbation, and then the nicely rolled vortices develop around
the discontinuity. The time of the individual snapshot is normalised by
$\tau_{KH}$, and shown at the upper--left corner of each frame.}
\label{2khi}
\end{figure*}

Figure \ref{2khi} shows the snapshots at different evolution times,
$t=0.5,\,1.0,\,1.5$ and $2.0\tau_{KH}$.
At $t=0.5\tau_{KH}$, the initial contact discontinuity is wiggling due to
the initial perturbation and the velocity shear.
There are nicely rolled vortices developing around the discontinuity in
the later snapshots.
A distortion of the vortices are observed in the snapshot
at $t=2.0\tau_{KH}$, and a mixing layer is expected to be formed
around the initial contact discontinuity.
Finally, the mixing layer will stop the KHI. Contrary to the standard
SPH, GSPH suffers from the unphysical force across the density gradient
much less than the standard SPH,
so it can describe the KHI in the different density layers.
Note that there isn't any kind of additional explicit
dissipation,
such as the artificial viscosity (or artificial conduction) in this
simulation.

\begin{figure*}
\includegraphics[width=8cm]{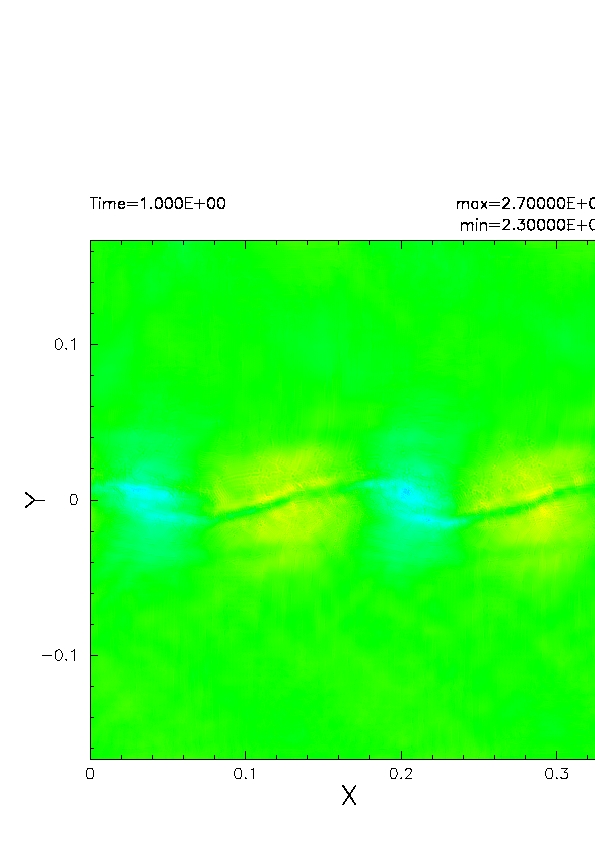}
\includegraphics[width=8cm]{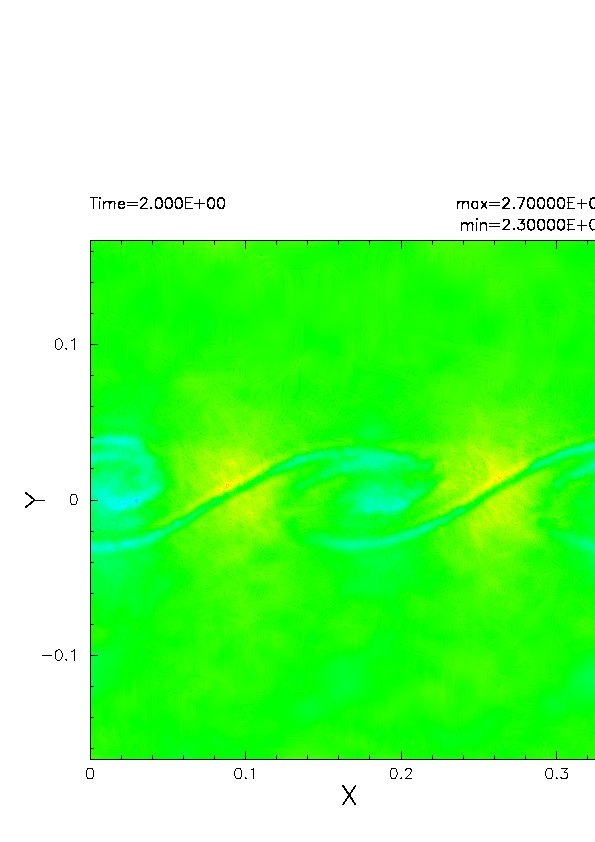}
\caption{Pressure distribution of the test shown in figure
\ref{2khi} at  $t=1.0$ and $2.0\tau_{KH}$. These pressure maps are less
noisy than the standard SPH results (see figure 6 of
\citet{Price2008a}). Note that \citet{Price2008a} has got a similar
result with the artificial conduction term.}
\label{2khi_pre}
\end{figure*}

Figure \ref{2khi_pre} is the pressure distribution at $t=1.0$ and
$2.0\tau_{KH}$, and shows a less noisy pattern while pressure blips are
observed across the contact discontinuity in the standard SPH result
(see figure 6 of \citet{Price2008a}). Note that \citet{Price2008a}
has got a similar pressure map with the artificial conduction as well.

\subsection{KHI in the two-layers $(\rho_{u}:\rho_{l} = 1:10)$}
The same KHI simulation presented in the previous section
has been performed again, but with a different density
contrast. The density contrast is much higher than the previous
simulation, and is set to 1:10.
The initial mach numbers are set to $0.2$ and $0.63$ in the
upper and lower layers, respectively. The total number of particles used
in this simulation is $\simeq 10^5$. The initial perturbation is the same
as the previous simulation.

\begin{figure*}
\includegraphics[width=8cm]{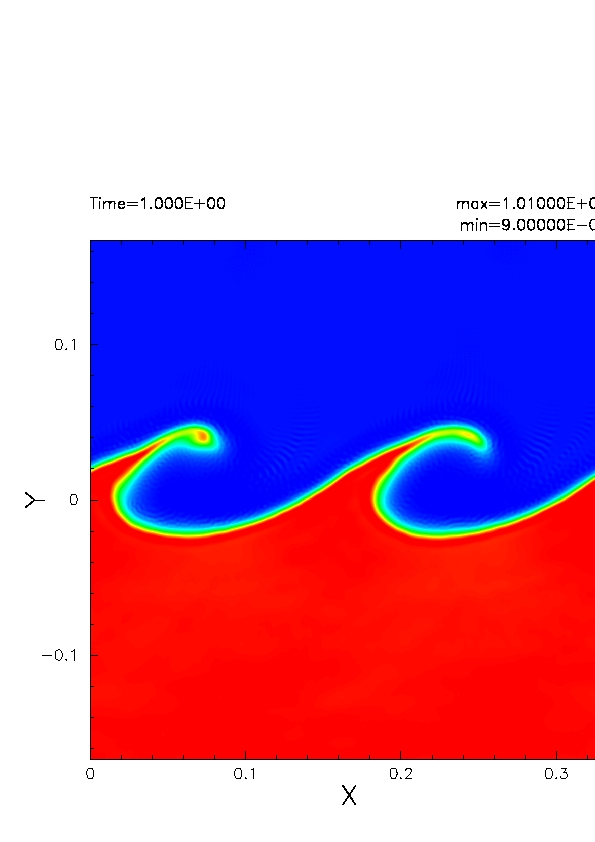}
\includegraphics[width=8cm]{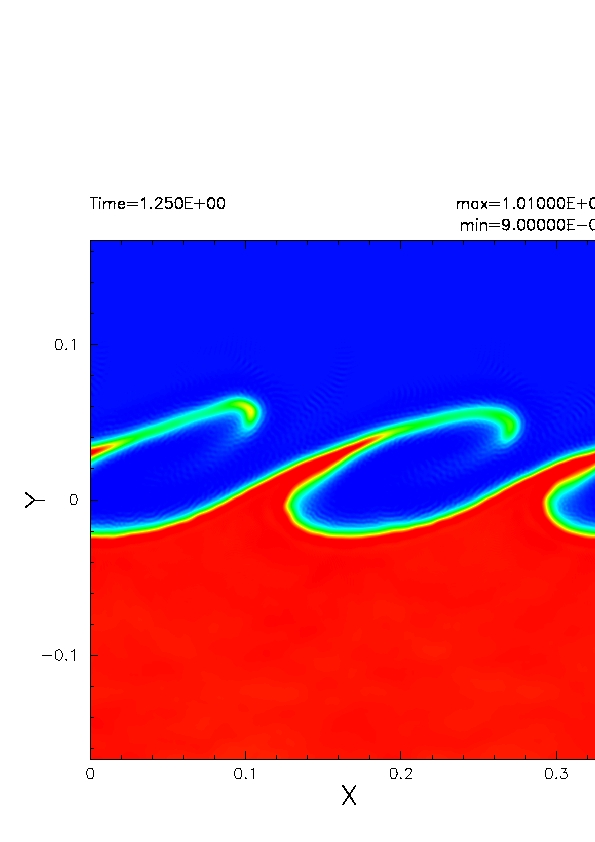}
\caption{The KHI simulation with a much higher density contrast.
The lower layer is 10 times denser than the upper layer.
The vortices develop around the discontinuity, and then
elongate in the long--term evolution. The elongation may be
due to the low resolution of the upper layer.}
\label{10khi}
\end{figure*}

Figure \ref{10khi} shows the results.
The two snapshots are at
$t=1.0$ and $1.25\tau_{KH}$, respectively.
The earlier stage than $1.0\tau_{KH}$ is very similar to the
lower density contrast case described in the previous section.
However, the vortices are not rolled but elongated in the later stage.
 
The reason of the elongation is not clear, but we guess that it may be
due to the poorer resolution of the upper layer
than the previous simulation \citep{Price2008a}.
GSPH (and also the standard SPH)
is a lagrangian method, so the numerical resolution depends on the
number density of particles. With a similar number of total particles,
the higher density contrast between the two
layers makes a poorer resolution of the lower density layer eventually.
Another possible reason of the vortex
elongation is the initial pressure.
We have used $2.5$ as the equilibrium pressure value in
this simulation, but different choice of the pressure value may change
the result. However, we'd like to emphasis that the KHI does happen
in this high density contrast case.

\subsection{KHI in the diagonal direction}
Contrary to grid--based Godunov schemes, in GSPH,
all interactions between the particles $i$ and $j$
reduce to a one--dimensional problem on the
line joining the two interacting particles even in a three--dimensional
problem. Therefore, a one--dimensional RPS is enough even in a
multi--dimensional GSPH code.
This is an advantage of GSPH than the grid--based
Godunov schemes, because there is no effective RPS
in multi--dimensional situation \citep{Monaghan1997a}. An
operator splitting method is essential in the grid--based
Godunov schemes to describe a multi--dimensional problem with
a one--dimensional RPS, but any
kind of geometrical splitting is not needed in GSPH.

\begin{figure*}
\includegraphics[width=8cm]{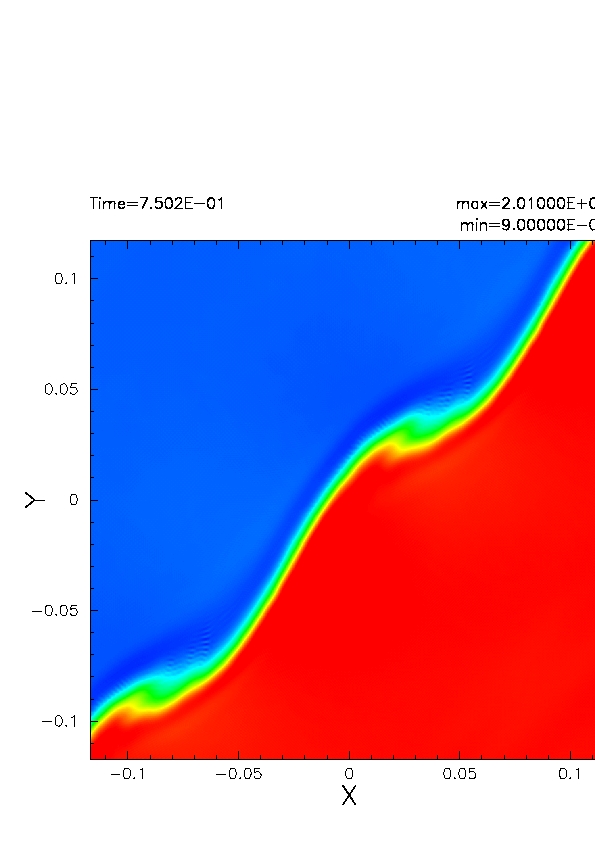}
\includegraphics[width=8cm]{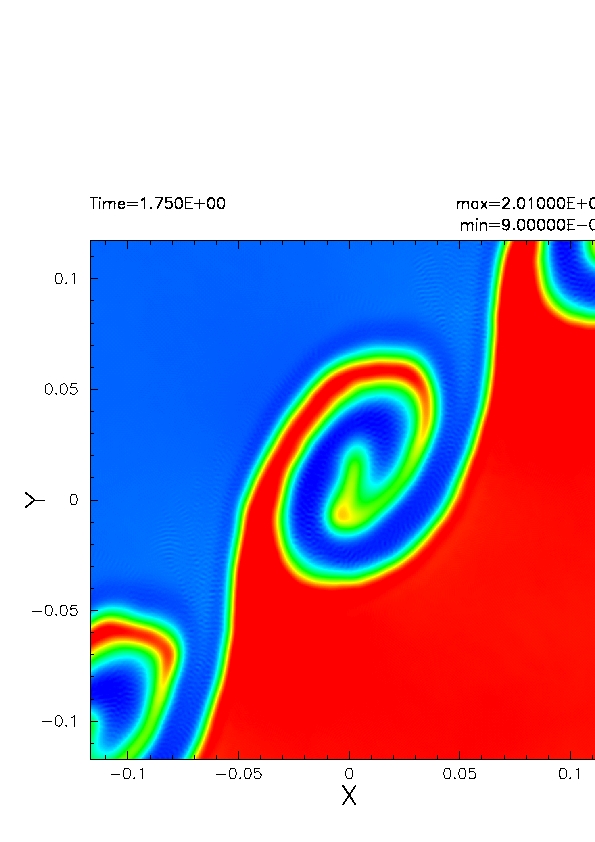}
\caption{The KHI developing in the diagonal direction.
This is essentially the same test presented in figure
\ref{2khi}, but the initial particle
distribution is rotated by $45^o$. Contrary to grid--based Godunov
methods, GSPH can describe a multi--dimensional
problem with a one--dimensional riemann problem solver,
so any kind of operator splitting is not needed.}
\label{2dia}
\end{figure*}

Figure \ref{2dia} shows the development of the KHI along the diagonal
direction. The density contrast is 1:2, and all initial conditions
are the same as the previous simulation described in section
\ref{sec_2khi} except the initial particle
distribution. The initial particle distribution is rotated by $45^o$.
One can see the well developed vortices along the diagonal direction in 
the figure.

\subsection{The blob test}
Interactions between dense blobs and strong blast waves are an
interesting subject in the context of the formation and evolution of
stars and galaxies
\citep{Murray1993a,Klein1994a,Jones1996a,Vietri1997a}.
If a dense blob is exposed to a strong blast wave (e.g. stellar wind
or supernova remnant),
the dense blob will be compressed due to the
blast wave initially, and destroyed finally.
The destruction
of the dense blob is initiated by the Rayleigh--Taylor and
Richtmyer-Meshkov instabilities \citep{Inogamov1999a},
and then enhanced by the KHI.
However, the instabilities or the combinations of instabilities
hardly happen in the standard SPH due to the unphysical force around the
front of the compressed blob, so the blob survives for a very long time
as it is compressed in the hot medium (A07).

We have performed the blob test with GSPH.
The calculation domain of the blob test is
$\left[-2,30\right]\times \left[-6,6\right]$ 
in code unit.
A dense blob is at the origin initially, and is surrounded by
the hot ambient medium moving in the $x$--direction.
The radius of the blob is 1, and the
density ratio between the ambient medium and the blob, $\chi$
is set to 10. The initial mach number of the ambient medium
is $5$. The numbers of particles to implement the blob and the
ambient medium are 7688 and 93139, respectively.
The initial configuration of the
particle distribution is the glass (A07).
The sound speed and the density of the ambient medium are set
to $1$. With
this initial condition, the cloud crushing time \citep{Klein1994a},
$\tau_{cc}$ is determined by
\begin{equation}
\tau_{cc} = \frac{r_b\sqrt{\chi}}{v_a},
\end{equation}
where $r_b$, and $v_a$ are the radius of the blob and the
velocity of the ambient medium, respectively.
\citet{Jones1996a} defined the ``bullet crushing time'', but the only
difference between $\tau_{cc}$ and the bullet crushing time
is a numerical factor $(=2)$, so we have used $\tau_{cc}$ as the
time unit in the
blob test. Finally, the KHI time scale (A07) of the blob test is defined by
\begin{equation}
\tau_{KHI,blob} = 1.6\times 2\tau_{cc}.
\end{equation}
$\tau_{cc}$ of this blob test is $0.63$.

\begin{figure*}
\includegraphics[width=5.7cm]{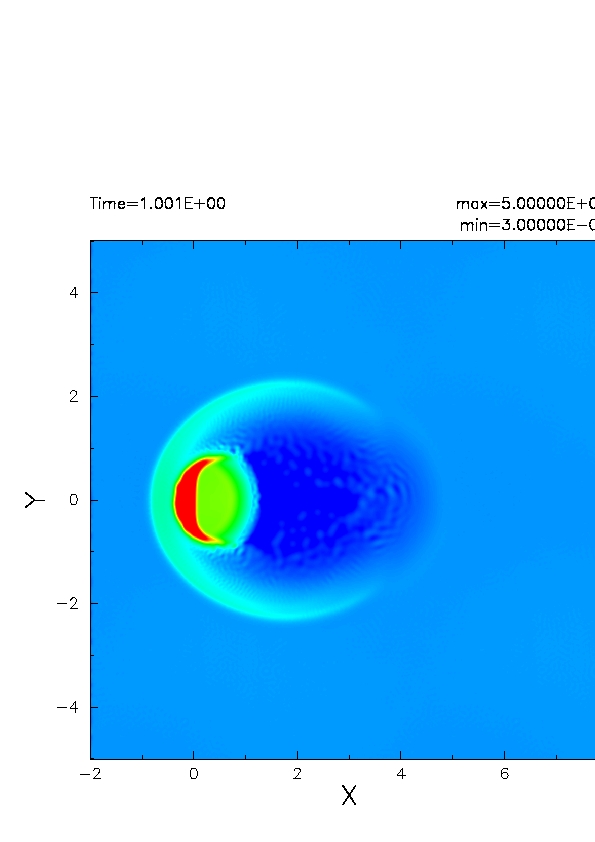}
\includegraphics[width=5.7cm]{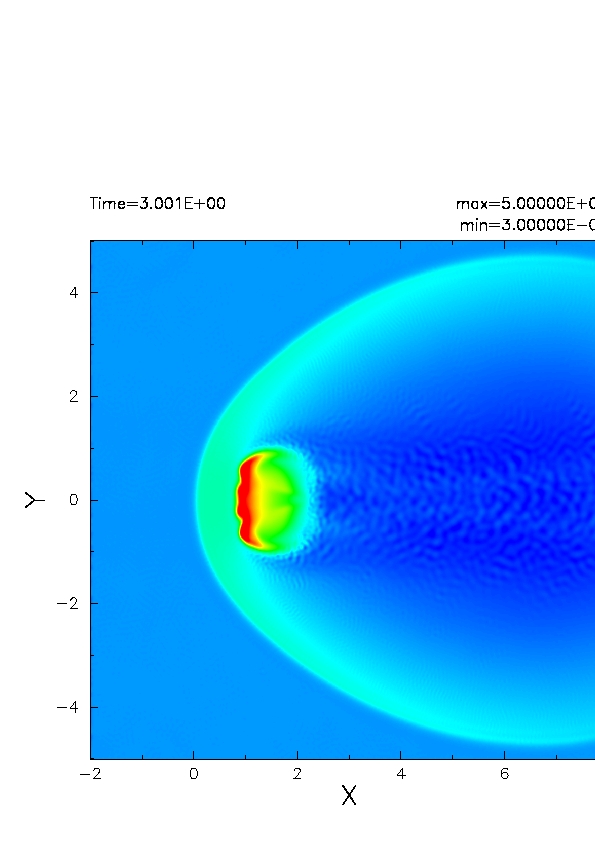}
\includegraphics[width=5.7cm]{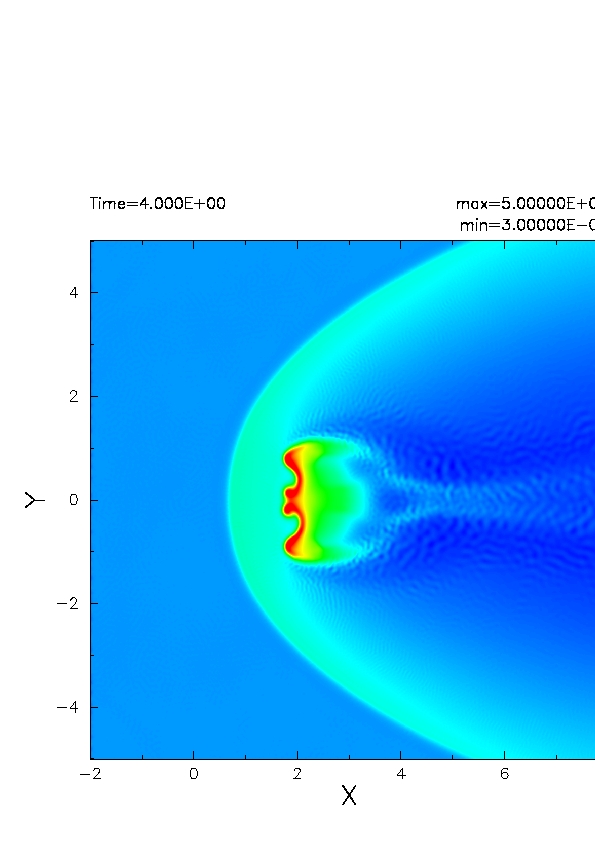}
\includegraphics[width=5.7cm]{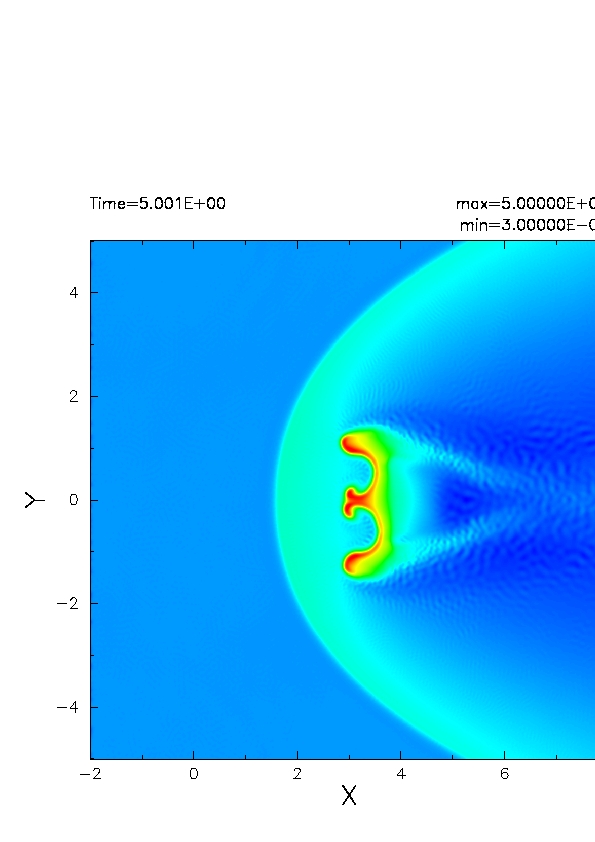}
\includegraphics[width=5.7cm]{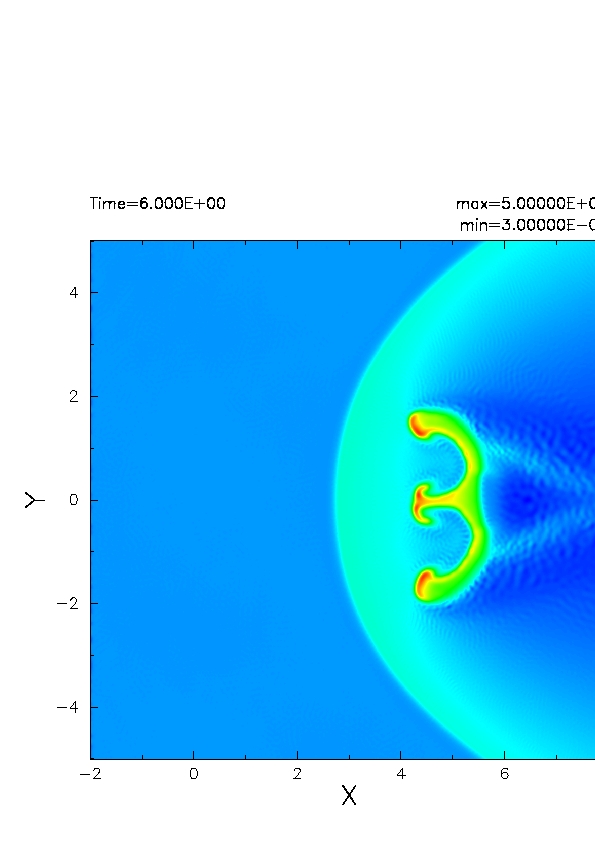}
\includegraphics[width=5.7cm]{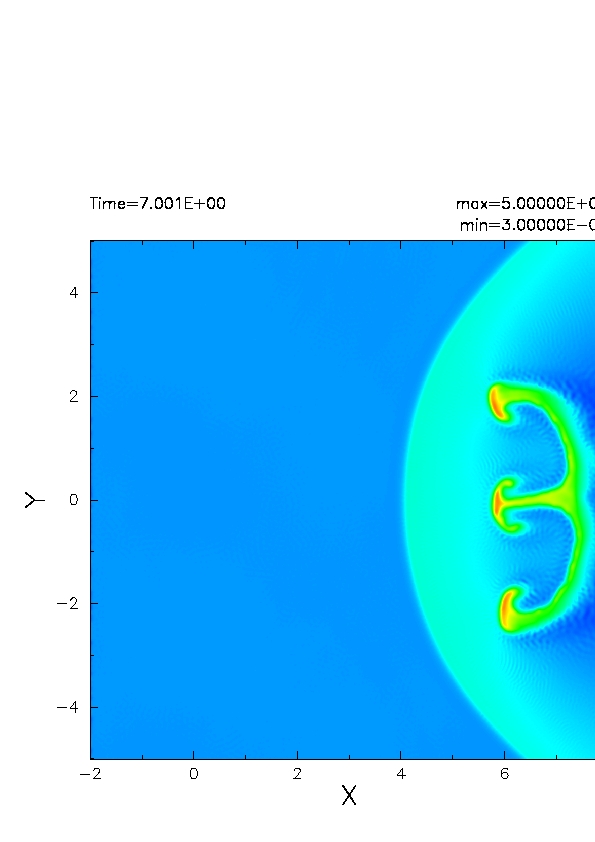}
\caption{Result of the blob test. Each snapshot is at
$t=1,\,3,\,4,\,5,\,6$ and $7\tau_{cc}$ from the top--left, and shows the
different stage of the blob evolution inside a hot moving ambient
medium. We have taken the square root of the column density to generate
the surface density color map.
The initial stage is the compression of the blob due to
the ram pressure of the ambient medium. One can see the bow shock formed
in front of the compressed blob. The ambient
medium enters into the bow shock is decelerated, and then eventually the
Rayleigh--Taylor and Richtmyer--Meshkov instabilities are triggered.
Three fingers begin to develop in front of the compressed blob
in the third snapshot,
and then the fingers are enhanced by the combinations of the
instabilities. A mushroom pattern develops at the head of the fingers.
This result coincides with the results of the grid--based
codes well. Note that this is a two dimensional simulation, while the
blob test of A07 has been performed in three dimensions.}
\label{blob}
\end{figure*}

Figure \ref{blob} is the result of the blob test. The early evolution
stage of the interaction is compression. The front of the
blob is compressed due to the ram pressure of the ambient medium. There
is an evaporation behind the blob. A bow
shock forms around the compressed blob, and then three fingers develop
due to the Rayleigh--Taylor and Richtmyer--Meshkov instabilities.
The fingers are enhanced by the KHI, so the mushroom pattern develops at
the head of the fingers \citep[e.g.][]{Yabe1991a}.
The result of the blob test performed by GSPH is similar to the
results of the grid--based code \citep[e.g.][]{Klein1994a}.

\section {Summary}
The standard SPH does not accurately
describe pressure gradient in the location with a
large density gradient, so
it shows the absence of the KHI in that situation.
This is due to the
inaccurate force calculation across the density gradient.
There is an unphysical force across the density gradient, and this
unphysical force
pushes the particles away from the the initial discontinuity
to make a gap and to damp the initial perturbation.
Therefore, the development of any
instability is suppressed at the density gradient.

The inaccurate force calculation
is due to the inconsistency of the standard SPH.
The particle approximation used in the derivation of the motion
equation of the standard SPH loses the $0^{th}$--order
consistency if particles are unevenly distributed.
One may use the Lagrangian function for the derivation of the
momentum equation, but 
the Lagrangian function of the standard SPH uses
the particle approximation
already, so the resulting momentum equation shows still
the unphysical force across a density gradient.

In order to solve the consistency problem of the standard SPH,
we have revisited the new formulation of
I02, called GSPH. With the kernel convolution, new momentum equation
is derived. We have proved that the momentum equation of GSPH
has linear consistency up to the accuracy with which the kernel
convolution integral can be calculated, leading to a much reduced
unphysical force across a density gradient in the pressure equilibrium.
The same momentum equation can be derived
using the new Lagrangian function
(I02). It is very similar to the Lagrangian function
of the standard SPH, but is
more accurate to the real fluid
Lagrangian function.

We have explained the geometrical meaning of the motion equation of
GSPH. It considers the host and neighbour particles as an extended body,
and uses the detailed information of the extended bodies.
In the standard SPH, the host particle is smoothed by the contributions
of neighbours, but the neighbours are considered as a point.

Two kinds of test have been performed to show the performance of GSPH.
One is the traditional KHI test in the two layers, and the other is the
blob test. In the two layer test, GSPH showed the development of the KHI
even with the very high density contrast.
The KHI developing along the diagonal direction has been performed also,
and a satisfying result has been obtained. In the blob test, GSPH can
describe the formation and evolution of the fingers due to the
instabilities and the combinations of instabilities in front of the
compressed blob. The blob test result of GSPH coincides with the results
of the grid--based codes.

In the standard SPH, not only the momentum equation, but also the energy
equation is inconsistent in the uneven particle distribution. We are
investigating the influence of the inconsistency on the energy equation,
and it is left for a following work.

\section*{Acknowledgements}
The authors thank to Oscar Agertz for the providing the details of the
tests performed in A07, and to Anthony Whitworth and Alexander Hobbs
for the useful discussion. We also appreciate for the very useful and
constructive comments of referee, Dr. Daniel Price. This work is
supported by the STFC Rolling grant.

\bibliographystyle{mn2e}

\bsp

\label{lastpage}

\end{document}